\documentclass[twocolumn,showpacs,aps,floatfix]{revtex4}

\usepackage{graphicx}

\def\lapproxeq{\lower .7ex\hbox{$\;\stackrel{\textstyle <}{\sim}\;$}}

\def\Ecut{E_c}
\def\mysection#1{\vspace*{2truemm}\noindent{\em #1}}

\begin{document}

\title{
Neutrinos: the Key to UHE Cosmic Rays
}
\author {David~Seckel \& Todor~Stanev}
\affiliation{
Bartol Research Institute, 
University of Delaware, Newark, DE 19716, USA
}
\widetext
{
\begin{abstract}

Observations of ultrahigh energy cosmic rays (UHECR) do not uniquely 
determine both the injection spectrum and the evolution model for UHECR 
sources - primarily because interactions during propagation obscure the 
early Universe from direct observation. Detection of neutrinos produced 
in those same interactions, coupled with UHECR results, would provide a 
full description of UHECR source properties.


\end{abstract}
 
}
\date{\today}

\maketitle


 Ultrahigh energy cosmic rays are one of the ten miracles of
 today's physics. Highest energy particles have long been
 suspected~\cite{Cocconi} to be extragalactic, because our Galaxy can 
 not magnetically contain, and respectively accelerate, them.
 Current experimental data on the highest energy cosmic rays
 show two severe problems:\\
 1.  It is equally difficult to explain their production with either 
traditional  astrophysical acceleration models~\cite{TA04}  or with 
exotic {\em top-down}~\cite{BhatS}  particle physics models. Acceleration 
models predict that UHECR  are charged nuclei, whereas {\em top-down} models 
predict them to be $\gamma$-rays and neutrinos.\\  
 2. Both nuclei and $\gamma$-rays  have small energy loss distances
 ${\cal{L}}_{loss} = (1/E)dE/dx$.
 Protons of 3$\times$10$^{20}$ eV lose their energy in propagation 
 on 23 Mpc. ${\cal{L}}_{loss}$ declines to less than 15 Mpc at
 higher energy. The $\gamma$-ray energy loss distance is less certain
 because of importance of the unknown isotropic radio
 background radiation. Reasonable estimates~\cite{ProthBier} 
 yield about 5 Mpc at energies between 10$^{19}$ and 10$^{21}$ eV.  
 Sources of UHECR then must be within several tens of Mpc, but none
 is identified.

 The solution of these problems is also affected by the
 current inconsistency in the results of the two major
 experimental groups~\cite{AGASA,HiRes}.
 HiRes data seem to show a GZK~\cite{GZK} cutoff, while AGASA's
 UHE cosmic ray spectrum can be explained only with the addition
 of nearby sources or top-down scenarios.  
 
\begin{figure}[thb] 
\centerline{\includegraphics[width=6.5cm]{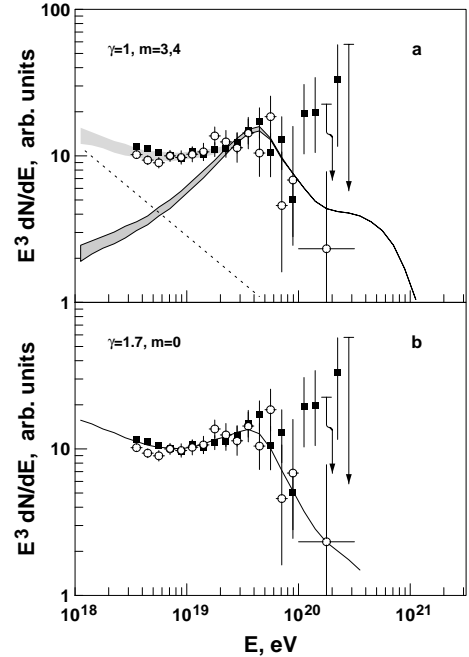}}
%
\caption{ a) A fit of UHECR spectrum with flat 
 extragalactic injection spectrum. Upper edge of shaded area is
 for \protect$m$=4, lower - \protect$m$=3. Galactic contribution is
 (dotted line) with \protect$\gamma$=2.6. b) Fitting with
 steep injection spectrum (\protect$\gamma$=1.7).
\label{ss:fig1}
}
\vspace*{-20pt}
\end{figure}

 Observations of the UHECR energy spectrum do not uniquely determine
 the extragalactic cosmic ray source distribution
 or the source spectrum even with high statistics - there are
 too many different ways to fit the spectrum. 
 Figure~\ref{ss:fig1} shows two extreme fits. The top panel 
 illustrates a fit with a flat E$^{-2}$ injection spectrum and 
 ($1+z$)$^{m}$($m$=3,4) cosmological evolution of the cosmic ray
 sources.   Such fits~\cite{WB,WW} can represent UHECR spectrum
 adequately if galactic cosmic rays extend 
 above 10$^{19}$ eV. Correspondingly, the chemical composition
 of cosmic rays contains a fraction of heavy nuclei up to that
 energy. The second cosmic ray knee~\cite{HiRes_comp} is where
 extragalactic cosmic rays prevail over the galactic ones.
 Fits with flat injection spectra require some cosmological
 evolution of the UHECR sources. Note, however, 
 the small effect of cosmological evolution on UHECR above
 10$^{18}$ eV - large redshifts do not contribute much.

 A different type of fit is illustrated in the bottom panel of
 Fig.~\ref{ss:fig1}. Extragalactic cosmic rays with injection
 spectrum E$^{-2.7}$ prevail down to 10$^{18}$ eV. These cosmic
 rays are expected to be protons and He nuclei. Generally
 injection spectra steeper than E$^{-2.5}$ can fit the observed
 spectra  without strong redshift evolution.
 The second knee is the result of $dE/dx$ due to pair
 production~\cite{BGG,BerGri}.
 The galactic cosmic ray spectrum extends to about 10$^{18}$ eV.
 The spectral shape of the extragalactic cosmic rays has to
 be flatter below 10$^{18}$ eV to avoid overproducing lower
 energy cosmic rays. One possible explanation is the 
 limited horizon of lower energy extragalactic cosmic rays 
 because of scattering in extragalactic magnetic
 fields~\cite{L04}.

\begin{figure}[thb] 
\centerline{\includegraphics[width=8.5cm]{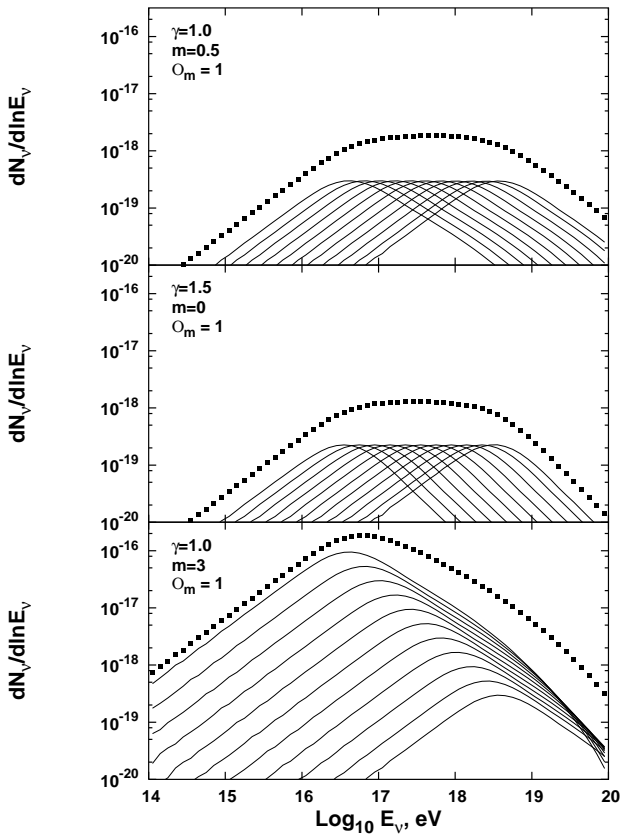}}
%
\caption{Neutrino production in three simplified models for the source 
spectrum and evolution of UHECR. To enhance the contrast between 
different models, the source luminosity is extended out to $1+ 
z_{max}=10$.
\label{ss:fig2}
}
\end{figure}

\mysection{Cosmogenic (GZK) neutrino production.} 
As shown in Fig. \ref{ss:fig1}, different UHE cosmic ray source models, 
with different luminosities at $10^{19}$~eV, produce spectra consistent 
with observations. The differences at lower energy may be disguised by 
contributions from the galaxy, or by propagation effects. The model 
degeneracy can be broken by considering the neutrino flux produced by the 
UHECR via their interactions on the cosmic micro-wave
background~\cite{BZ69,S73}, the so-called GZK neutrino flux.

To make the point, we consider a set of simple power law models in a 
matter dominated cosmology, with a homogeneous distribution of UHECR 
sources. On cosmological time scales, the UHECR interact quickly with the 
CMB, so we model the production of GZK neutrinos as instantaneous with 
particle injection. We use the neutrino yields of Ref.~\cite{ESS}. The 
neutrino flux at Earth due to GZK production is
\begin{equation} 
\label{ss:eq1}
E_\nu \frac{d\Phi}{dE_\nu}(E_\nu) = \int dt d\epsilon_p 
\frac{d\Gamma}{d\epsilon_p} 
E_\nu \frac{dy}{dE_\nu}(E_\nu,\epsilon_p,t) 
\end{equation}
where $\Gamma$ is the injection rate of UHECR and $y$ is the neutrino yield 
per proton injected with energy $\epsilon_p$, and $E_\nu$ is the neutrino 
energy today.

Equation \ref{ss:eq1} can be put into a more convenient form by defining 
$q=1+z$ and integrating over redshift. For a matter dominated cosmology, 
$H_0 dt = -q^{-3/2} d(\ln q)$, where $H_0$ is the present day 
Hubble parameter. As a function of red shift, we model the UHECR 
injection rate as
\begin{equation} 
\frac{d\Gamma}{d\epsilon_p}=q^m \epsilon_p^{-(1+\gamma)} A
\end{equation}
where $\gamma$ is the integral spectral index of the source, $m$ 
describes the evolution of the co-moving source density, and $A$ is 
 chosen to normalize the emissivity of UHECR sources to their
 energy density in the present Universe.
 We use  ${\cal L}_{CR} = 4.5 \times {\rm 10}^{44}$ erg/Mpc$^3$/yr
 as estimated  by Waxman~\cite{W95}.
 The neutrino yield can also be scaled with redshift.
 Defining the present day yield as $\frac{dy_0}{dE_\nu}(E_\nu, E_p)$,
 the yield from a previous epoch is
\begin{equation} 
E_\nu\frac{dy}{dE_\nu}(E_\nu, \epsilon_p,z) 
  = E_\nu \frac{dy_0}{dE_\nu}(q^2 E_\nu, q \epsilon_p)
\end{equation} 
The factors of $q$ are due to the redshift of neutrino energy from 
production to the current epoch, and to the lowering of the reaction 
threshold due to the increased CMB temperature in the early Universe. 
Let $Y_0$ be the integrated yield from injecting 
an $E_p^{-(1+\gamma)}$ spectrum today:
$E_\nu\frac{dY_{0\gamma}}{dE_\nu}(E_\nu) = \int 
dE_p E_p^{-(1+\gamma)} 
E_\nu\frac{dy_0}{dE_\nu}(E_\nu, E_p)$.
Then the integral yield at any other redshift 
is given by 
\begin{equation}
\int d\epsilon_p \epsilon_p^{-(1+\gamma)} 
E_\nu\frac{dy}{dE_\nu}(E_\nu,\epsilon_p,t)
= q^{\gamma} E_\nu\frac{dY_{0\gamma}}{dE_\nu}(q^2 E_\nu)
\end{equation}
With these definitions, the GZK production integral can be expressed as 
an integral over redshift,
\begin{equation} 
E_\nu \frac{d\Phi}{dE_\nu}(E_\nu) = \frac{A}{H_0} 
\int_0^{q_{max}} d(\ln q) q^{(m+\gamma-\frac{3}{2})} 
E_\nu\frac{dY_{0\gamma}}{dE_\nu}(q^2 E_\nu)
\end{equation}

Written as an integral over $\ln q$, it is straightforward to see which 
epoch dominates the neutrino flux. If $m+\gamma=1.5$ then all redshift 
intervals contribute with comparable importance. This is illustrated by 
the top panel of Fig.~\ref{ss:fig2}, where $m=0.5, \gamma=1$. The thin 
lines represent the contribution to the neutrino flux from epochs spaced 
equally in $\ln q$, and of equal width in $d(\ln q)$. The peak 
contribution for each interval occurs at an energy $E_{pk}(q)$, which 
scales with redshift as $E_{pk}(q)=E_{pk}(0)/q^2$.
The sum of the thin lines, 
back to a red shift of $1+z_{max}=10$, gives the dotted curve - the 
predicted GZK neutrino flux for the model. The integrated flux is flat, 
with a width in $E_\nu$ of order $(1+z_{max})^2$.  

Similarly, the middle panel illustrates a second model with equal 
contribution per epoch, although here there is no 
evolution ($m=0$) and the spectrum is correspondingly steeper ($\gamma 
 =1.5$). The total cosmogenic neutrino flux is slightly lower because of
 the smaller number of interacting protons. 
 The larger value of $\gamma$ is only evident through the 
different slope of the high energy part of the neutrino spectrum. In 
contrast, if $m+\gamma>1.5$ the neutrino flux is dominated by past 
epochs, as illustrated in the bottom panel of Fig.~\ref{ss:fig2}. This 
corresponds to the situation for a flat source model with significant 
 evolution, $m=3, \gamma=1$. Note that this illustrative model is not
 realistic as the $(1+z)^3$ evolution continues to 1+$z_{max}$ = 10. 


\begin{figure}[thb] 
\centerline{\includegraphics[width=8.5cm]{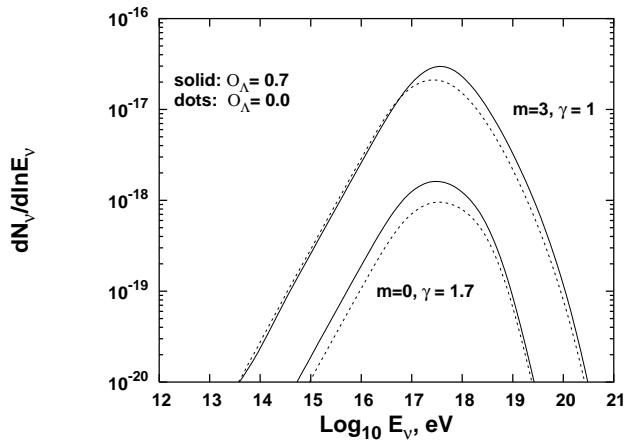}}
%
\caption{Neutrino production for the UHECR models shown in Fig. 
\ref{ss:fig1}. The injection spectrum is cutoff at $\Ecut=3 \times 
10^{21}$~GeV and the source evolution breaks at $z_{max}=3$. Model 
results depend mildly on the cosmological model, $\Omega_\Lambda = 
0.0~{\rm or}~0.7$.
\label{ss:fig3}
}
\end{figure}

\mysection{Discussion.} The scaling analysis illustrates the combined 
importance of the UHECR spectrum and cosmological evolution in estimating 
the neutrino flux due to GZK production. For more realistic models, it is 
appropriate to consider other effects, such as a cosmological constant, a 
cutoff to the cosmic ray injection spectrum, and a cutoff or break in 
scale to the source evolution model. These effects alter the details of 
GZK production, but do not change the overall picture. Accordingly, 
Fig.~\ref{ss:fig3} shows an order of magnitude difference in the GZK 
production for the two extreme UHECR models discussed in 
Fig.~\ref{ss:fig1}. Apart from the lower value of $z_{max}$, the steep 
source model with no evolution ($\gamma=1.7, m=0$) is similar to the 
middle panel of Fig. \ref{ss:fig2}. The model with flat spectrum and  
cosmological evolution ($\gamma=1, m=3$) predicts an order of 
magnitude larger flux.

This brings us to the main point of this paper. The AUGER observatory, 
under construction, will measure the intensity and spectrum of UHECR. 
Above the GZK cutoff these are observations of the current Universe. 
Below the GZK cutoff it may be difficult to separate the old 
intergalactic cosmic rays from a population of young cosmic rays which 
originate within our own galaxy.
 Complementary to AUGER, experiments are being designed and
 constructed (ANITA, IceCube, Mediterranean km$^3$, RICE), which
 may confirm the existence of GZK neutrinos. A next generation
 of experiments (EUSO, OWL, SalSA,  X-RICE) is  being planned
 which would provide sufficient statistics (10-100 GZK events
 per yr) to complement and expand the AUGER observations.
 Successful completion of one such experiment would be an important
 step toward understanding the sources of the highest energy particles
 in the Universe. 

It is often argued that neutrino astronomy has value in that neutrinos 
allow observation of the interior of objects, whereas photons only allow 
observation of the surface. Solar neutrinos are an example, confirming 
theoretical models of the thermonuclear furnace in the sun, by direct 
observation of the core. The current discussion is similar. The 
``surface'' is the interaction distance for super-GZK UHECR. The 
``interior'' is the early Universe, where the evolution of UHECR sources 
can be directly observed through the GZK neutrino flux.

\mysection{Acknowledgments.} 
This research is supported in part by NASA Grant NAG5-10919.
TS is also supported by the US Department of Energy contract
DE-FG02 91ER 40626.

\bibliographystyle{prsty}

\end{document}